\documentclass[twocolumn,prb,showpacs]{revtex4-1}
\usepackage[colorlinks,bookmarks=false,citecolor=blue,linkcolor=red,urlcolor=blue]{hyperref}
\usepackage{graphicx,epstopdf,color}
\usepackage{amsfonts}
\usepackage{amsmath,amssymb,mathrsfs}
\usepackage{bm}
\usepackage{ucs} 
\usepackage[utf8x]{inputenc} 
\usepackage[T1]{fontenc}

\newcommand{\ex}[1]{\exp\hspace{-2pt}\left({#1}\right)}
\newcommand{\exbig}[1]{\exp\hspace{-.5pt}\bigl({#1}\bigr)}
\newcommand{\exBig}[1]{\exp\hspace{0pt}\Bigl({#1}\Bigr)}

\newcommand{\ext}[1]{\exp\hspace{-1.5pt}\left(\textstyle{#1}\right)}

\newcommand{\tauy}{\text{\text{Im}}(\tau)}

\newcommand{\pz}{\partial_{z}}

\newcommand{\be}{{\bar\eta}}

\newcommand{\pbe}{\partial_{\bar\eta}}
\renewcommand{\tt}{\tilde t}
\newcommand{\thetau}[1]{
  \vartheta_{\frac{1}{2}\hspace{-.6pt},\hspace{-.6pt}\frac{1}{2}}
  \!( #1|\tau)}
\newcommand{\thetaubig}[1]{
  \vartheta_{\frac{1}{2}\hspace{-.6pt},\hspace{-.6pt}\frac{1}{2}}
  \!\big( #1\big|\tau\big)}
\newcommand{\thetauBig}[1]{
  \vartheta_{\frac{1}{2}\hspace{-.6pt},\hspace{-.6pt}\frac{1}{2}}
  \!\Big( #1\Big|\tau\Big)}
\newcommand{\thetaubigg}[1]{
  \vartheta_{\frac{1}{2}\hspace{-.6pt},\hspace{-.6pt}\frac{1}{2}}
  \!\bigg( #1\bigg|\tau\bigg)}

\newcommand{\thembtaubig}[1]{
  \vartheta_{\frac{1}{2}\hspace{-.6pt},\hspace{-.6pt}\frac{1}{2}}
  \!\big( #1|-\!\bar\tau\big)}
\newcommand{\thembtauBig}[1]{
  \vartheta_{\frac{1}{2}\hspace{-.6pt},\hspace{-.6pt}\frac{1}{2}}
  \!\Big( #1|-\!\bar\tau\Big)}

\newcommand{\bs}[1]{\boldsymbol{#1}}

\newcommand{\comm}[2]{\left[#1,#2\right]}
\newcommand{\bigcomm}[2]{\bigl[#1,#2\bigr]}

\renewcommand{\i}{\text{i}}

\newcommand{\z}{\text{z}}

\newcommand{\abs}[1]{\left|#1\right|}

\newcommand{\parder}[2]{\frac{\partial #1}{\partial #2}}

\renewcommand{\a}{\alpha}

\newlength{\ytlength}
\def\ie{{i.e.},\ }
\def\eg{{e.g.}\ }

\allowdisplaybreaks[1]
\begin{document}
\title{Laughlin states and their quasi-particle excitations on the
  torus}
\author{Martin Greiter}
\author{Vera Schnells}
\author{Ronny Thomale}
\affiliation{Institute for Theoretical Physics, University of
  W\"urzburg, Am Hubland, 97074 W\"urzburg, Germany}
\pagestyle{plain}
\begin{abstract}
  We provide a full derivation of Laughlin's Jastrow-type wave
  functions for quantized Hall states subject to periodic boundary
  conditions using an operator formalism.  The construction includes
  the quasi-hole and the technically more challenging quasi-electron
  excitation, which was left as an open problem in the classic paper
  by Haldane and Rezayi [Phys.~Rev.~B {\bf 31}, 2529 (1985)].
\end{abstract}




\maketitle

\section{Introduction}

Laughlin's theory of the fractional quantized Hall
effect~\cite{laughlin83prl1395} has been arguably one of the most
influential development in theoretical condensed matter physics in the
past decades.  It has been generalized to other rational filling
odd-denominator Landau level fractions through a hierarchy of
fractionally quantized Hall
states~\cite{haldane83prl605,halperin84prl1583,greiter94plb48,Jain07},
but also to even denominator filling fractions through inclusion of
p-wave pairing
correlations~\cite{moore-91npb362,greiter-91prl3205,greiter-92npb567}.
The fractionally charged quasi-particle excitations of the
odd-denominator states are the first realizations of particles obeying
fractional statistics~\cite{wilczek90,stern08ap204} in two space
dimensions (Abelian 2D anyons), while the quasi-particles in the
paired odd-denominator states and their generalizations obey
non-Abelian
statistics~\cite{moore-91npb362,read-00prb10267,ivanov01prl268,stern10n187}
(non-Abelian 2D anyons).  The concept of non-Abelian statistics, which
now plays a preeminent role in the field of quantum
computation~\cite{kitaev03ap2,nayak-08rmp1083}, was actually
discovered in paired Hall states~\cite{moore-91npb362}, where it is
understood in terms of unpaired Majorana
fermions~\cite{majorana37nc171,read-00prb10267} at the quasi-particle
vortex cores.  Wave functions similar to the Laughlin and Read--Rezayi
series~\cite{read-99prb8084} of quantized Hall states describe the
exact ground states of an integrable spin one-half chain
model~\cite{haldane88prl635,shastry88prl639} and its generalizations
to critical models for higher spin
chains~\cite{Greiter11,thomale-12prb195149}.  These models have led us
to understand Abelian~\cite{haldane91prl937,greiter09prb064409} and
non-Abelian anyons in one dimension~\cite{Greiter11}.  The concept of
topological order~\cite{wen90ijmpb239,Wen04,Bernevig13} was introduced
in the context of the Abelian chiral spin
liquid~\cite{kalmeyer-87prl2095,kalmeyer-89prb11879,schroeter-07prl097202,thomale-09prb104406},
which is a spin liquid in which bosonic spin flip operators are
described by a Laughlin--Jastrow-type Hall wave function.

Laughlin's wave functions~\cite{laughlin83prl1395,laughlin84ss163}, as
originally proposed, describe circular droplets of fractionally
quantized Hall fluids in the open plane, the geometry Laughlin had
used for early numerical work.  For the definite confirmation of its
correctness, Haldane eliminated the boundary by reformulating the
theory on a sphere in a magnetic monopole
field~\cite{haldane83prl605,fano-86prb2670,greiter11prb115129}, and
then showed numerically that the energy gap in the excitation spectrum
remained intact as the Hamiltonian is continuously varied from
screened Coulomb interactions to the parent Hamiltonian for the
Laughlin states.  The third important geometry is the torus or
periodic boundary conditions, 
the only among them with a non-zero genus ($g=1$), and hence the only
one in which the topological degeneracies originally discovered by
Haldane and Rezayi~\cite{haldane-85prb2529} are observable.  These
degeneracies have then been interpreted as a manifestation of
topological order, a new concept introduced by
Wen~\cite{wen90ijmpb239,Wen04} to describe order in quantum Hall and
spin liquid states.  All of this has been understood in the early
years, and has now been well established for decades.

There is one small, technical aspect in the formulation of Laughlin's
theory, however, which was not resolved by the masters. 
Wave functions for the Laughlin states and the quasi-hole excitations
were given by Haldane and Rezayi~\cite{haldane-85prb2529}, but
formulating the technically more challenging quasi-electron excitation
was left as an open problem.  The formulation of ground states and
quasi-holes was subsequently generalized to include Pfaffian states on
the torus~\cite{greiter-92npb577}, but then again, the quasi-electron
excitations were not addressed.  Furthermore, an elegant formulation
of quantized Hall wave functions in terms of conformal correlators was
developed~\cite{moore-91npb362,hansson-07prb075347,bergholtz-07prl256803}.
While quantum Hall ground states were generalized to the torus for
composite fermions and the Haldane-Halperin hierarchy through
conformal field
theory~\cite{hermanns-08prb125321,hermanns13prb235128,fremling-14prb125303},
an explicit formulation of Laughlin's quasi-electron wave function has
so far only been reported in the open
plane~\cite{hansson-09prb165330}.

Presumably, there are two reasons for this.  First, it is highly
non-trivial from a technical point of view.  Haldane and
Rezayi~\cite{haldane-85prb2529} devoted a paragraph to the problem in
their classic paper, but the form of an Ansatz they suggest is already
inconsistent with the results we present below.  Second, it does not
seem to be of tremendous importance, as our understanding of
fractionally quantized Hall liquids is rather complete, and the number
of questions we might wish to address at this stage is limited.  To
mention two possible applications, one could compare Jain's composite
fermion quasi-electron (and hierarchy) wave functions to Laughlin's
construction in a plane without boundaries.  This would provide a
complementary view to similar investigations on the
sphere~\cite{jeon-03prb165346}, where the result reflects the
different clustering properties of the Jack polynomials describing
both states~\cite{bernevig-09prl066802}.  
Furthermore, modular invariance arising in the context of the
calculation of Hall viscosity~\cite{read09prb045308} might be
investigated with the toroidal Laughlin quasi-particle wave functions
at hand, and shed further light on similarities and differences to
alternative approaches~\cite{fremling-14prb125303}.

Viewed from a broader perspective, however, it is best not to judge
the importance of solutions to problems before we have obtained them.
In solving a problem, and in solving a hard problem in particular, we
may learn something which might turn out useful in a different context
later on.  With this in mind, we now revisit Laughlin's theory on the
torus, and present a complete formulation.  

The paper is organized as follows.  In sections ~\ref{sec:landau}
and~\ref{sec:single}, we review the ladder operator formalism for
Landau level quantization, quasi-periodic boundary conditions, and
single-particle states on the torus in symmetric gauge.  Following
Haldane and Rezayi~\cite{haldane-85prb2529}, we then construct
Laughlin's wave functions for the ground states and quasi-hole
excitations of fractionally quantized Hall liquids on the torus in
sections~\ref{sec:ground} and~\ref{sec:hole}.  In
section~\ref{sec:electron}, we finally construct Laughlin's
quasi-electron wave function to the torus.  In the conclusion~\ref{sec:conclusion}, we compare our result for the
quasi-electron wave function to the quasi-hole wave function of
Haldane and Rezayi, and speculate why the problem of constructing the
quasi-electron has not been solved previously.

\section{Landau levels, ladder operators, and magnetic translations}
\label{sec:landau}   

To describe the dynamics of charged particles (\eg spin-polarized
electrons) in a two-dimensional plane subject to a perpendicular
magnetic field $\bs{B}=-B\bs{e}_\z $, it is convenient to introduce
complex particle coordinates $z=x+\text{i}y$ and $\bar
z=x-\text{i}y$~\cite{landau30zp629,Arovas86}.  The associated derivative
operators are
\begin{align}
  \label{eq:partialz}
  \partial_{z}
  =\frac{1}{2}\left(\partial_{x}-\text{i}\partial_{y}\right),\
  \partial_{\bar z}
  =\frac{1}{2}\left(\partial_{x}+\text{i}\partial_{y}\right).
\end{align}
Note that hermitian conjugation yields a $-$ sign,
\begin{align}
  \label{eq:hconjparderz}
  \left(\partial_{z}\right)^\dagger
  =-\partial_{\bar z}.
\end{align}
The single particle Hamiltonian is obtained by minimally coupling the
gauge field to the canonical momentum,
\begin{align}
  \label{eq:SinglePartHam1}
  H=\frac{1}{2M}\left(\bs{p}+\frac{e}{c}\bs{A}\right)^2,
\end{align}
where $M$ is the mass of the particle and $e>0$.  For our purposes,
and in particular for the formulation of the quasi-electron
excitations on the torus, it is more convenient to work in symmetric
gauge $\bs{A}=\frac{1}{2}B\,\bs{r}\times\bs{e}_\z$, rather than in the
Landau gauge used by Haldane and Rezayi~\cite{haldane-85prb2529}.

With the definition of the magnetic length
$l=\sqrt{{\hbar c}/{eB}}$
and the ladder
operators~\cite{macdonald84prb3550,girvin-83prb4506,Arovas86}$^,$%
\footnote{We have not been able to find out who introduced the ladder
  operators for Landau levels in the plane.  The energy eigenfunctions
  were known since Landau~\cite{landau30zp629}.
  MacDonald~\cite{macdonald84prb3550} used the ladder operators in
  1984, but neither gave nor took credit.  Girvin and
  Jach~\cite{girvin-83prb4506} were aware of two independent ladders a
  year earlier, but neither spelled out the formalism, nor pointed to
  references.  It appears that the community had been aware of them,
  but not aware of who introduced them.  Clear and complete
  presentation can be found in Arovas~\cite{Arovas86} or
  Greiter~\cite{Greiter11}.}  
describing the cyclotron variables
\begin{align}
  \label{eq:aladder1}
  a=\sqrt{2}l
  \left(\partial_{\bar z}+\frac{1}{4l^2}z\right),\
  a^\dagger=\sqrt{2}l
  \left(-\partial_{z}+\frac{1}{4l^2}\bar z\right),  
\end{align}
which obey $\comm{a}{a^\dagger}=1$,
we may rewrite \eqref{eq:SinglePartHam1} as
\begin{align}
  \label{eq:SinglePartHam3}
  H=\hbar\omega_\text{c}\left(a^\dagger a+\frac{1}{2}\right),
\end{align}
where $\omega_\text{c}=eB/Mc$ is the cyclotron frequency.  The kinetic
energy of charged particles in a perpendicular magnetic field is hence
quantized like a harmonic oscillator (Landau level quantization).  The
lowest Landau level (LLL) consists of those states annihilated by $a$.
We introduce a second set of ladder operators describing the 
guiding center variables,
\begin{align}
  \label{eq:bladder1}
  b=\sqrt{2}l
  \left(\partial_{z}+\frac{1}{4l^2}\bar z\right),\
  b^\dagger=\sqrt{2}l
  \left(-\partial_{\bar z}+\frac{1}{4l^2}z\right).  
\end{align}
They likewise obey $\comm{b}{b^\dagger}=1$ and commute with the
cyclotron ladder operators, 
\begin{align}
  \label{eq:Commab}
    \comm{a^{\phantom{\dagger}}\!\!}{b}=\comm{a}{b^\dagger}=0.
\end{align}
For our purposes, it will be convenient to write the ladder
operators as 
\begin{align}
  \label{eq:abLadder}
  (a,b)&=\sqrt{2}l\hspace{2pt} 
   e^{-S}\left(\partial_{\bar z},\partial_{z}\right) e^{+S},
   \\[0.5\baselineskip] \label{eq:adagbdagLadder}
  (a^\dagger,b^\dagger)&=\sqrt{2}l\hspace{2pt} 
   e^{-S}\left(-\partial_{z}+\frac{1}{2l^2}\bar z,
   -\partial_{\bar z}+\frac{1}{2l^2} z \right) e^{+S},
\end{align}
where $S\equiv|z|^2/4l^2$.    
Since the canonical angular momentum 
\begin{align}
  \label{eq:SinglePartL1}
  \bs{L}=\bs{r}\times\bs{p}
  =\hbar \left(b^\dagger b - a^\dagger a\right)\bs{e}_\z 
\end{align}
commutes with the Hamiltonian \eqref{eq:SinglePartHam3}, we can use it
to classify the vastly degenerate states within each Landau level.  A
complete, orthonormal set of basis states is 
\begin{align}
  \label{eq:Phinm}
  \phi_{n,m}(z)
  =\frac{1}{\sqrt{n!}}\frac{1}{\sqrt{m!}}(a^\dagger)^n (b^\dagger)^m 
  \frac{1}{\sqrt{2\pi} l} e^{-S},
\end{align}
where we omitted $\bar z$ from the argument of the wave functions as a
choice of convention.
Here $n+1$ denotes the Landau level index, and $m-n$ the angular
momentum around the origin.  With the particle position given by
$z=\sqrt{2}l\left(a+b^{\dagger}\right)$, we find for the states
\eqref{eq:Phinm} in the LLL 
\begin{align}
  \label{eq:LLLm}
  \phi_{0,m}(z)
   &=   \frac{1}{\sqrt{2^{m+1}\pi m!}\; l^{m+1}}\hspace{2pt} z^m e^{-S}.
\end{align}
These states is describe narrow rings centered around the origin, 
with the radius determined by
\begin{equation*}
  \label{eq:rm}
  \parder{}{r}\abs{\phi_{0,m}(r)}^2\biggl|_{r=r_m}\overset{!}{=}0\biggr.,
\end{equation*}
which yields $r_m=\sqrt{2m}\,l$.  Since there are also $m$ states
inside the ring, the areal degeneracy is
\begin{align}
  \label{eq:ArealDeg}
  \frac{\text{number of states}}{\text{area}}
  =\frac{m}{\pi r_m^2}=\frac{1}{2\pi l^2},
\end{align}
The magnetic flux required for each state,
\begin{equation*}
  2\pi l^2 B =\frac{2\pi \hbar c}{e}=\Phi_0,
\end{equation*}
is hence given by the Dirac flux quantum.  

We can write the most general single particle state in the LLL as
\begin{align}
  \label{eq:SinPartLLL}
  \psi(z)=e^{-S} g(z),
\end{align}
where $g(z)$ is an analytic function of $z$.  

Since the Hamiltonian \eqref{eq:SinglePartHam3} does not commute with
translations, but only with translations supplemented by gauge
transformations, we introduce the magnetic translation operator
\begin{align}
  \label{eq:t}
  t(\xi)\equiv\ex{\frac{1}{\sqrt{2}l}(\xi b-\bar\xi b^\dagger)}.
\end{align}
It obviously commutes with \eqref{eq:SinglePartHam3}, but 
Baker--Hausdorff,
\begin{align}
  \nonumber
  e^{x+y} = e^x\, e^y\, e^{-\frac{1}{2}\comm{x}{y}}\ \ \text{for}\
  \comm{\comm{x}{y}}{x}=\comm{\comm{x}{y}}{y}=0,
\end{align}
implies
\begin{align}
  \label{eq:t1t2}
  t(\xi_1) t(\xi_2)=t(\xi_1+\xi_2) 
  \ex{\frac{1}{4l^2}(\bar\xi_1 \xi_2 -\xi_1\bar\xi_2)},
\end{align}
and $\bar\xi_1 \xi_2 -\xi_1\bar\xi_2=2\i(\bs{r}_1\times\bs{r}_2)_\z$,
magnetic translations along different directions $t(\xi_1)$ and
$t(\xi_2)$ commute only if the area spanned by $\xi_1$ and $\xi_2$ in
the plane contains an integer number $N_\phi$ of magnetic Dirac flux
quanta, \ie $(\bs{r}_1\times\bs{r}_2)_\z=2\pi l^2 N_\phi$.  This
condition reflects Dirac's quantization condition for magnetic
monopoles, which implies that the magnetic flux through any closed
surface is given by an integer number of Dirac flux quanta.

\section{Single particle states on the torus}
\label{sec:single}

We impose periodic boundary conditions (PBCs) by
\begin{align}
  \label{eq:PBCs}
  t(\xi_\a)\hspace{1pt} \psi(z)
  =e^{\i\phi_\a}\psi(z),\quad\text{for}\ \a=1,\tau,
\end{align}
where $\xi_1$ and $\xi_\tau$ are two nonparallel displacements in the
complex plane, and $\phi_1$ and $\phi_\tau$ are boundary phases.
Since the boundary conditions require that $t(\xi_1)$ and $t(\xi_\tau)$
commute, the parallelogram spanned by  $\xi_1$ and $\xi_\tau$ must
contain an integer number of flux quanta, which we call $N_\phi$.

Due of the magnetic field, the wave functions subject to
\eqref{eq:PBCs} are not strictly periodic, but only quasi-periodic,
\begin{align}
  \label{eq:psiQuasiperiodic}
  \psi(z+\xi_\a)=
    \ex{-\frac{1}{4l^2}(\xi_\a\bar z - \bar \xi_\a z)} e^{\i\phi_\a}\psi(z).
\end{align}
We set the principal displacements \mbox{$\xi_1=1$} and
$\xi_\tau=\tau$, with $\tauy>0$, and call the region
bounded by the four points $z=\frac{1}{2}(\pm1\pm\tau)$ the principal
region.  This fixes the magnetic length according to 
\begin{align}
  \label{eq:lNphi}
  2\pi l^2 N_\phi = \tauy.
\end{align}
For our purposes, it is most convenient to write the eigenstates as
\begin{align}
  \label{eq:psiz}
  \psi(z)=e^{-S}\,f(z)\,e^{{z^2}/{4l^2}}.
\end{align}
These states are related by a simple gauge transformation from Landau
to symmetric gauge to the eigenstates used by Haldane and
Rezayi~\cite{haldane-85prb2529} (see Eqn.\ (1) of their paper).
It is further convenient to introduce the operator
\begin{align}
  \label{eq:ttilde}
  \tt(\xi)&\equiv e^{+S} t(\xi) e^{-S}
  \nonumber \\[0.2\baselineskip] 
  &=\ex{\xi\partial_{z}+\bar\xi\partial_{\bar z}
    -\frac{1}{2l^2}\bar\xi z}
  \nonumber \\[0.2\baselineskip] 
  &=\ex{-\frac{1}{4l^2}\bar\xi (2z+\xi)}\ext{\xi\partial_{z}}
\end{align}
where we omitted the term involving $\partial_{\bar z}$.  (This is
possible because $\tilde t(\xi)$ acts throughout this article only on
a function of $z$, not $\bar z$.)  With
\begin{align}
  \label{eq:ttilde1ONeTOzTO2}
  \tt(1)\, e^{{z^2}/{4l^2}}&=e^{{z^2}/{4l^2}},
  \\[0.3\baselineskip] \label{eq:ttildetauONeTOzTO2}
  \tt(\tau)\, e^{{z^2}/{4l^2}}&=e^{{z^2}/{4l^2}}\,
  e^{i\pi N_\phi (2z+\tau)},
\end{align}
the PBCs \eqref{eq:PBCs} for $\psi(z)$ as specified by \eqref{eq:psiz}
imply
\begin{align}
  \label{eq:fRatio1}
  \frac{f(z+1)}{f(z)}&=e^{\i\phi_1},
  \\[0.3\baselineskip]\label{eq:fRatiotau} 
  \frac{f(z+\tau)}{f(z)}&=e^{-\i\pi N_\phi (2z+\tau)} e^{\i\phi_\tau}.
\end{align}
As Haldane and Rezayi~\cite{haldane-85prb2529} have pointed out, 
\eqref{eq:fRatio1} and \eqref{eq:fRatiotau} can be used to calculate
the numbers of zeros of $f(z)$ in the principal region (PR).  With the
theorem of residues, we obtain
\begin{align}
  \label{eq:Nzerosf}
  \frac{1}{2\pi\i}\oint_{\text{PR}}\! dz\frac{f'(z)}{f(z)}
  \hspace{-69pt}&\hspace{69pt}
  =\frac{1}{2\pi\i}\oint_{\text{PR}}\! d(\ln f(z))
  \nonumber\\[0.3\baselineskip]\nonumber
  &=\frac{1}{2\pi\i}\left[
  \int_{-\frac{1}{2}+\frac{\tau}{2}}^{-\frac{1}{2}-\frac{\tau}{2}}\hspace{-3pt}
  d\!\left[\ln\frac{f(z)}{f(z+1)}\right]
  +\int_{-\frac{1}{2}-\frac{\tau}{2}}^{\frac{1}{2}-\frac{\tau}{2}}\hspace{-3pt}
  d\!\left[\ln\frac{f(z)}{f(z+\tau)}\right]\right]
  \nonumber\\[0.3\baselineskip]
  &=\frac{1}{2\pi\i}
  \int_{-\frac{1}{2}-\frac{\tau}{2}}^{\frac{1}{2}-\frac{\tau}{2}}\hspace{-3pt}
  d\!\left(i\pi N_\phi (2z+\tau)-\i\phi_\tau\right)=N_\phi
\end{align}
The function $f(z)$, and hence also the wave function $\psi(z)$,
has hence as many zeros as there are Dirac flux quanta in the
principal region.  

The most general form for $f(z)$ is hence
\begin{align}
  \label{eq:fSingle}
  f(z)=e^{\i kz} \prod_{\nu=1}^{N_\phi} \thetau{z-z_\nu},
\end{align}
where $k$ is a real parameter $|k|\le\pi N_\phi$, all the zeros $z_\nu$
are located in the principal region, and
$\thetau{\hspace{1pt}z\hspace{1pt}}$ is the odd Jacobi theta
function~\cite{Mumford83}.  The theta functions are defined in general
for $a,b=0,\frac{1}{2}$ by
\begin{align}
  \label{eq:thetadef}
  \vartheta_{a,b}(z|\tau ) = \sum_{n=-\infty }^{+\infty }\,e^{\pi
    \i(n+a)^2\tau}\,e^{2\pi \i(n+a)(z+b)}
\end{align}
and satisfy the quasi-periodicity relations
\begin{align}
  \label{eq:theta+1}
  \vartheta_{a,b}(z+1|\tau)&=e^{2\pi\i a}\,
  \vartheta_{a,b}(\hspace{1pt}z\hspace{1pt}|\tau ),
  \\[0.3\baselineskip]\label{eq:theta+tau}
  \vartheta_{a,b}(z+\tau|\tau)&=e^{-\pi\i\tau}\, e^{-2\pi\i (z+b)}\, 
  \vartheta_{a,b}(\hspace{1pt}z\hspace{1pt}|\tau ).
\end{align}
The latter formula implies
\begin{align}
  \label{eq:theta+ntau}
  \vartheta_{a,b}(z+n\tau|\tau)&=e^{-\pi\i\tau n^2}\, e^{-2\pi\i n(z+b)}\, 
  \vartheta_{a,b}(\hspace{1pt}z\hspace{1pt}|\tau ).
\end{align}
Substitution of \eqref{eq:fSingle} into \eqref{eq:fRatio1} and
\eqref{eq:fRatiotau} yields that $k$ and the zeros $z_\nu$ are subject
to the boundary conditions
\begin{align}
  \label{eq:bc1}
  &(-1)^{N_\phi}\ex{\i k} = e^{\i\phi_1},
  \\[0.3\baselineskip]\label{eq:bctau}
  &(-1)^{N_\phi}\ex{\i k\tau}\ex{2\pi\i\sum_{\nu=1}^{N_\phi} z_\nu} = e^{\i\phi_\tau}.
\end{align}
For fixed values of the boundary phases $\phi_1$ and $\phi_\tau$,
\eqref{eq:bc1} and \eqref{eq:bctau} possess a total of $N_\phi^2$
solutions for $k$ and $\sum_\nu z_\nu$; no further restrictions for
the allowed choices of the individual $z_\nu$ result.  Of all these
distinct solutions, however, only $N_\phi$ yield linearly independent
states.  This is physically obvious, as there are only $N_\phi$
independent states in the LLL.  It can also be seen from a simple
mathematical consideration, as we will elaborate now.
 
The abstract property we require of $f(z)$ is to be analytic in $z$
and have exactly $N_\phi$ zeros in the principal region. Given one
such solution $f(z)$, its ratio ${\tilde f(z)}/f(z)$ with any other
solution $\tilde f(z)$ is a meromorphic truly periodic function on the
torus, with at most simple poles at $N_\phi$ prescribed points (namely
the zeroes of $f(z)$).  It is a general theorem of complex function
theory (a special case of the Riemann-Roch theorem) that the space of
such functions is $N_\phi$ dimensional, including the constant
function~\cite{JonesSingerman87}.  Hence there are only $N_\phi$
independent states.

\section{Laughlin's ground states}
\label{sec:ground}

Generalizing the formalism developed in the previous section, we write
the most general $N$ particle state in the LLL as
\begin{align}
  \label{eq:psi[z]}
  \psi[z]=e^{-S}\, f[z]\,\prod_{i=1}^N e^{{z_i^2}/{4l^2}},
\end{align}
where $[z]\equiv (z_1,z_2,\ldots ,z_N)$, 
\begin{align}
  \label{eq:S[z]}
  e^{-S} =\prod_{i=1}^N e^{-{|z_i|^2}/{4l^2}},
\end{align}
and $f(z_1,z_2,\ldots,z_N)$ is a completely antisymmetric (symmetric)
function of the $z_i$'s for fermions (bosons).  It is subject to the
PBCs
\begin{align}
  \label{eq:iPBCs}
  t_i(\xi_\a)\hspace{1pt} \psi[z]
  =e^{\i\phi_\a}\psi[z],\quad\text{for all}\ i,\ \a=1,\tau,
\end{align}
where 
\begin{align}
  \label{eq:ti}
  t_i(\xi)\equiv\ex{\frac{1}{\sqrt{2}l}(\xi b_i-\bar\xi b_i^\dagger)}.
\end{align}
effects a magnetic translation of particle $z_i$.  As in the previous
section, the PBCs imply
\begin{align}
  \label{eq:f1Ratio1}
  \frac{f(z_1+1,z_2,\ldots,z_N)}{f[z]}
  &=e^{\i\phi_1},
  \\[0.3\baselineskip]\label{eq:f1Ratiotau} 
  \frac{f(z_1+\tau,z_2,\ldots,z_N)}{f[z]}
  &=e^{-\i\pi N_\phi (2z_1+\tau)} e^{\i\phi_\tau}.
\end{align}

The characteristic feature of Laughlin's Jastrow-type states at Landau
level filling fraction $\nu=1/m$ is that a test particle $z_1$ sees
exactly $m$ zeros at the locations of all the other particles
$z_2,\ldots,z_N$.  With $N_\phi=m N$, this requirement fixes the
location of all but $m$ zeros in each of the coordinates.  Following
Haldane and Rezayi~\cite{haldane-85prb2529}, we are led to try an
Ansatz
\begin{align}
  \label{eq:deff}
  f[z]=F(Z)\,\prod_{i<j}^N\thetau{z_i-z_j}^m,
\end{align}
where 
\begin{align}
  \label{eq:Z}
  Z\equiv\sum_{i=1}^N z_i
\end{align}
is the center-of-mass coordinate.
Substitution of \eqref{eq:deff} into \eqref{eq:f1Ratio1} and
\eqref{eq:f1Ratiotau} yields
\begin{align}
  \label{eq:FRatio1}
  \frac{F(Z+1)}{F(Z)}&=(-1)^{N_\phi-m}\,e^{\i\phi_1},
  \\[0.3\baselineskip]\label{eq:FRatiotau} 
  \frac{F(Z+\tau)}{F(Z)}
  &=(-1)^{N_\phi-m}\,e^{-\i\pi m (2Z+\tau)}\,e^{\i\phi_\tau}.
\end{align}
These equations are extremely similar to \eqref{eq:fRatio1} and
\eqref{eq:fRatiotau}, and imply via 
\begin{align}
  \label{eq:NzerosF}
  \frac{1}{2\pi\i}\oint_{\text{PR}}\! d(\ln F(Z))=m
\end{align}
that $F(Z)$ has exactly $m$ zeros in the principal region.

In analogy to \eqref{eq:fSingle}, the most general solution for the
single particle states
\begin{align}
  \label{eq:FSingle}
  F(Z)=e^{\i KZ} \prod_{\nu=1}^{m} \thetau{Z-Z_\nu},
\end{align}
where $K$ is a real parameter $|K|\le\pi m$, and all the zeros $Z_\nu$
are located in the principal region.  Substitution of
\eqref{eq:FSingle} into \eqref{eq:FRatio1} and \eqref{eq:FRatiotau}
yields that $K$ and the center-of-mass zeros $Z_\nu$ are subject to
the boundary conditions
\begin{align}
  \label{eq:Fbc1}
  &(-1)^{N_\phi}\ex{\i K} = e^{\i\phi_1},
  \\[0.3\baselineskip]\label{eq:Fbctau}
  &(-1)^{N_\phi}\ex{\i K\tau}\ex{2\pi\i\sum_{\nu=1}^{m} Z_\nu} = e^{\i\phi_\tau}.
\end{align}
For fixed values of the boundary phases $\phi_1$ and $\phi_\tau$,
\eqref{eq:Fbc1} and \eqref{eq:Fbctau} possess a total of $m^2$
solutions for $k$ and $\sum_\nu z_\nu$.  As above, only $m$ of these
solutions yield linearly independent states.  We can see this either
from the equivalence of the problem to \eqref{eq:bc1} and
\eqref{eq:bctau}, or directly from the mathematical argument
indicated there.
 
The $m$-fold degeneracy of the Laughlin $1/m$ state on a closed 
surface with genus one is a topological quantum number which 
characterizes the topological order of the state. 

\section{Quasi-hole excitations}
\label{sec:hole}

The generalization of this formalism to include a quasi-hole
excitation at position $\eta$ in the complex plane follows without
incident.  While in many applications it is sufficient to view $\eta$
as a parameter, there are some which require to view it as a
dynamical variable.  An example of an application of the latter type
is the hierarchy of quantized Hall states, where the quasi-particle
excitations themselves condense into Laughlin--Jastrow-type fluids~\cite{haldane83prl605,halperin84prl1583,greiter94plb48}.

We hence set up the formalism as general as possible, and write 
the wave function with one quasi-hole as
\begin{align}
  \label{eq:psie[z]}
  \psi_\eta[z]=e^{-S}\,f_\eta[z]\;e^{{\eta^2}/{4ml^2}}\prod_{i=1}^N e^{{z_i^2}/{4l^2}},
\end{align}
where
\begin{align}
  \label{eq:Seta[z]}
  e^{-S} =e^{-{|\eta|^2}/{4ml^2}}\prod_{i=1}^N e^{-{|z_i|^2}/{4l^2}},
\end{align}
and $N_\phi=mN+1$.  Since the quasi-hole carries a fraction of $1/m$
of an electron charge, the effective magnetic length for the
quasi-holes is $\sqrt{m}$ times the magnetic length $l$ seen by the
electrons.  In analogy to the ground state, the PBCs \eqref{eq:iPBCs}
for \eqref{eq:psie[z]} imply the conditions \eqref{eq:f1Ratio1} and
\eqref{eq:f1Ratiotau} for $f_\eta[z]$.

Since the quasi-hole excitation amounts to an isolated zero seen by
all the particles, we consider an Ansatz
\begin{align}
  \label{eq:deffeta}
  f_\eta[z]=F_\eta(Z)\,
  \prod_{i=1}^N\thetau{z_i-\eta}\,\prod_{i<j}^N\thetau{z_i-z_j}^m,
\end{align}
where the center-of-mass coordinate now includes the quasi-hole,
\begin{align}
  \label{eq:Feta}
  F_\eta(Z)=e^{\i K\left(Z+\eta/m\right)} 
  \prod_{\nu=1}^{m} \thetauBig{Z+\frac{\eta}{m}-Z_\nu},
\end{align}
With $Z$ once again given by \eqref{eq:Z}, we find that
\eqref{eq:f1Ratio1} and \eqref{eq:f1Ratiotau} for $f_\eta[z]$ are
satisfied if K and the center-of-mass zeros $Z_\nu$ are once again
subject to the boundary conditions \eqref{eq:Fbc1} and
\eqref{eq:Fbctau}.

Regarding the quasi-periodicity in the quasi-hole coordinate $\eta$,
we begin by introducing ladder operators for the quasi-hole
coordinate,
\begin{align}
  \label{eq:cLadder}  
  c&=\sqrt{2m}l\hspace{2pt} e^{-S}\hspace{1pt}\partial_{\eta} e^{+S},
  \\[0.5\baselineskip] \label{eq:cdagLadder}
  c^\dagger&=\sqrt{2m}l\hspace{2pt} 
   e^{-S}\hspace{-1pt} 
   \left(-\partial_{\bar\eta}+\frac{1}{2ml^2}\eta\right) e^{+S},
\end{align}
where $S$ is given by \eqref{eq:Seta[z]}.  (Note that for the purposes
of defining the operators, we could have just as well taken
$S=|\eta|^2/4ml^2$, as the terms depending on the $z_i$'s cancel in
\eqref{eq:cLadder} and \eqref{eq:cdagLadder}.)  $c$ and $c^\dagger$
obey the commutation relation $\comm{c}{c^\dagger}=1$ and trivially
commute with $a$, $a^\dagger$, $b$, and $b^\dagger$.  We define the magnetic
translation operator for the quasi-hole coordinate
\begin{align}
  \label{eq:te}
  t_\eta(\xi)\equiv\ex{\frac{1}{\sqrt{2m}l}(\xi c-\bar\xi c^\dagger)}.
\end{align}
Since
\begin{align}
  \label{eq:t1t2eta}
  \hspace{-4pt}t_\eta(\xi_1) t_\eta(\xi_2)=t_\eta(\xi_1+\xi_2) 
  \ex{\frac{1}{4ml^2}(\bar\xi_1 \xi_2 -\xi_1\bar\xi_2)},
\end{align}
magnetic quasi-hole translations along different directions
$t_\eta(\xi_1)$ and $t_\eta(\xi_2)$ commute only if the area spanned
by $\xi_1$ and $\xi_2$ in the plane contains $m$ times an integer
number $N_\phi$ of magnetic Dirac flux quanta (or equivalently,
an integer number of ``quasi-hole flux quanta'' $m\Phi_0$, where 
$\Phi_0$ is the Dirac flux quantum).

From both these consideration and \eqref{eq:Feta}, we conclude that
in general, we do not expect the quasi-hole wave function to be
quasi-periodic under translations by $1$ and $\tau$, but only under
translations by $m$ and $m\tau$.  As a side note, we can achieve
quasi-periodicity by quasi-hole translations of either $1$ and $m\tau$
or $m$ and $\tau$, if we arrange the center-of-mass zeros with equal
spacing along one of the meridians of the torus,
\begin{align}
  \label{eq:ZinLine}
  Z_\nu=Z_1+\frac{(\nu-1)}{m}
  \quad\text{or}\quad 
  Z_\nu=Z_1+\frac{(\nu-1)\tau}{m}. 
\end{align}
From \eqref{eq:deffeta} with \eqref{eq:Feta},
we can readily derive
\begin{align}
  \nonumber 
  \frac{f_{\eta+m}[z]}{f_\eta[z]}
  &=(-1)^{m(N+1)} e^{\i K}, 
  \\[0.3\baselineskip]
  \frac{f_{\eta+m\tau}[z]}{f_\eta[z]}
  &=(-1)^{m(N+1)}e^{\i K\tau} e^{-\i\pi N_\phi (2\eta +m\tau)}
  \nonumber\\\nonumber 
  &\hspace{9pt}\cdot \ex{2\pi\i\sum_{\nu=1}^{m} Z_\nu}. 
\end{align}
In analogy to our elaborations in Section \ref{sec:single}, we
introduce an operator
\begin{align}
  \label{eq:tetatilde}
  \tt_\eta(\xi)&\equiv e^{+S} t_\eta(\xi) e^{-S}
  \nonumber \\[0.2\baselineskip] 
  &=\ex{\xi\partial_{\eta}+\bar\xi\partial_{\bar\eta}
    -\frac{1}{2ml^2}\bar\xi\eta}
  \nonumber \\[0.2\baselineskip] 
  &=\ex{-\frac{1}{4ml^2}\bar\xi (2\eta+\xi)}\ext{\xi\partial_{\eta}},
\end{align}
where we omitted the term involving $\partial_{\bar\eta}$ (knowing that
it will only act on functions of $\eta$, not $\bar\eta$).  With 
\begin{align}
  \label{eq:ttilde1ONe^eta^2}
  \tt_\eta(m)\, e^{{\eta^2}\hspace{-2pt}/{4ml^2}}&=e^{{\eta^2}\hspace{-2pt}/{4ml^2}},
  \\[0.3\baselineskip] \label{eq:ttildetauONe^eta^2}
  \tt_\eta(m\tau)\, e^{{\eta^2}\hspace{-2pt}/{4ml^2}}&=
  e^{i\pi N_\phi (2\eta+m\tau)}\,e^{{\eta^2}\hspace{-2pt}/{4ml^2}},
\end{align}
we find that
\begin{align}
  t_\eta(m)\hspace{1pt}\psi_\eta[z]
  &=(-1)^{N_\phi+m-1} e^{\i K}\hspace{1pt}\psi_\eta[z],
  \nonumber\\[0.3\baselineskip] \nonumber
  t_\eta(m\tau)\hspace{1pt}\psi_\eta[z]
  &=(-1)^{N_\phi+m-1}e^{\i K\tau}
  \ex{2\pi\i\sum_{\nu=1}^{m} Z_\nu}\psi_\eta[z].
\end{align}
With the boundary conditions \eqref{eq:Fbc1} and \eqref{eq:Fbctau},
this implies that
\begin{align}
\label{eq:tetaPBC}
  t_\eta(m\xi_\a)\hspace{1pt}\psi_\eta[z]
  &=e^{\i \phi_\a}(-1)^{m-1}\hspace{1pt}\psi_\eta[z],\quad\text{for}\ \a=1,\tau,
\end{align}
with $\xi_1=1$ and $\xi_\tau=\tau$.  
For fermions, $m$ is odd and $(-1)^{m-1}=1$.  As we (magnetically)
translate the quasi-hole with charge $e^*=e/m$ $m$ times around one of
the meridians of the torus, the ground state wave function acquires
the same phase as it acquires when we translate an electron with
charge $-e$ once around.

The example of arranging the center-of-mass zeros according to
\eqref{eq:ZinLine} lends itself well to illustrate the connection
between translations of the quasi-hole once around one of the
meridians of the torus and the topological degeneracy of the Laughlin
states on higher genus surfaces.  As both cases mentioned above are
analogous, we only elaborate the first one, where
\begin{align}
  \label{eq:Z1inLine}
  Z_\nu=Z_1+\frac{(\nu-1)}{m}.
\end{align}
If we (magnetically) translate the quasi-hole by $1$, the
center-of-mass zeros transform according to
\begin{align}
  \label{eq:ZnuTrans1}
  Z_\nu\to Z_{\nu-1}\quad \text{for}\ \nu>1,\quad Z_1\to Z_m. 
\end{align}
Magnetic translation of the quasi-hole by 1 will hence only yield
a phase in the wave function,
\begin{align}
\label{eq:tetaPBCline1}
  t_\eta(1)\hspace{1pt}\psi_\eta[z]
  &=(-1)^{N+1} e^{\i K/m}\hspace{1pt}\psi_\eta[z].
\end{align}
If we magnetically translate the quasi-hole via $t_\eta(\tau)$ by
$\tau$, however, all the center-of-mass zeros will by shifted,
\begin{align}
  \label{eq:ZnuTranstau}
  Z_\nu\to Z_{\nu}-\frac{\tau}{m}\quad \text{for}\ \nu.
\end{align}
We hence obtain a different, topologically degenerate ground state.
From the real part of boundary condition \eqref{eq:Fbctau}, we see
that \eqref{eq:ZnuTranstau} implies $K\to K+2\pi$, and hence via
\eqref{eq:tetaPBCline1} that the new state is orthogonal to the
original one.  After $m$ magnetic translations by $\tau$, we finally
recover the original state modulo a phase, as specified by
\eqref{eq:tetaPBC}.

Even though we might have clarified some details and increased the
accessibility, all of the results presented so far have been
understood by Haldane and Rezayi~\cite{haldane-85prb2529}, as well as
the community at large.  The wave functions for the quasi-electron
excitations on the torus we derive in the following section, however,
have to our knowledge not been obtained previously.

\section{Quasi-electron excitations}
\label{sec:electron}

Before we dive into the details of how to construct Laughlin's
quasi-electron wave function on the torus, let us briefly recall the
construction in the plane.  Laughlin obtains the quasi-hole by
inserting one Dirac flux quantum adiabatically in the direction of the
background magnetic field at a position $\eta$.  In the process, all
the electrons in the liquid acquire a shift of $\hbar$ in the
canonical angular momentum around $\eta$, and the ground state evolves
into
\begin{equation}
  \label{eq:psiQH}
  \psi_\eta[z]
  =\prod^N_{i=1}(z_i-\eta) \prod^N_{i<j}(z_i-z_j)^m\, e^{-S}.
\end{equation}
As $m$
quasi-holes at $\eta$ would constitute a true hole with charge $+e$
in the liquid, the charge of the quasi-hole is $+e/m$.

The quasi-electron, \ie the antiparticle of the quasi-hole, has charge
$-e/m$ and is created by inserting the flux adiabatically in the
opposite direction, thus lowering the angular momentum around some
position $\eta$ by $\hbar$, or alternatively, by removing one of the
zeros from the wave function.  To accomplish this formally, we first
rewrite \eqref{eq:psiQH} in terms of ladder operators:
\begin{equation}
  \label{eq:psiQHladder}
  \psi_\eta[z]
  =\prod^N_{i=1}\left(\sqrt{2}l\hspace{1pt}b^\dagger_i-\eta\right)
  \prod^N_{i<j}(z_i-z_j)^m \, e^{-S}.
\end{equation}
The insertion of a flux quantum in the opposite direction, or the
lowering of the angular momentum around $\eta$, will then correspond to
the Hermitian conjugate operation.  Laughlin~\cite{laughlin84ss163}
hence proposed for the quasi-electron wave function in the plane
\footnote{As a side remark, numerical work on the sphere shows that
  the quasi-electron wave function can be improved energetically by
  acting only on a Jastrow factor squared with the derivatives.
  This is likely to hold for the torus as well.}
\begin{align}
  \label{eq:psiQE}
  \psi_\be[z]
  &=\prod^N_{i=1}\left(\sqrt{2}l\hspace{1pt}b_i-\be\right)
  \prod^N_{i<j}(z_i-z_j)^m \, e^{-S}
  \nonumber\\[0.3\baselineskip]
  &=e^{-S}\,
  \prod^N_{i=1}\left(2l^2\partial_{z_i}-\be\right)
  \prod^N_{i<j}(z_i-z_j)^m.
\end{align}
Conducting the same operation with the quasi-hole factors 
in \eqref{eq:deffeta} we obtain with \eqref{eq:thetadef},
\begin{align}
  \label{eq:qeFactor}
  \left[\thetauBig{\sqrt{2}l\hspace{1pt}b^\dagger_i -\eta}\right]^\dagger
  =\thembtauBig{\sqrt{2}l\hspace{1pt}b_i -\be}.
\end{align}
These considerations suggest that for the wave function with a
quasi-electron at $\be$ on the torus, we may consider an Ansatz of the
form
\begin{align}
  \label{eq:psibe[z]}
  \psi_\be[z]=e^{-S}\,f_{\be,\pbe }[z,\pz ]\,e^{R}
\end{align}
where $S$ is still given by \eqref{eq:Seta[z]}, $f$ depends now on
derivatives of the coordinates $z_i$ and $\be$ as well, $N_\phi=mN-1$,
and we have defined
\begin{align}
  \label{eq:R}
  e^{R}\equiv e^{{\be^2}/{4ml^2}}\prod_{i=1}^N e^{{z_i^2}/{4l^2}}.
\end{align}
The challenge is now to identify the functions $f_{\be,\pbe }[z,\pz]$
such that $\psi_\be[z]$ is quasi-periodic under translations of
any $z_i$ around either meridians of the torus, and of $\be$ under
$m$ such translations in either direction.

Regarding the $z_i$'s, the PBCs are still given by \eqref{eq:iPBCs}
with \eqref{eq:ti}.  Since $f_{\be,\pbe }[z,\pz]$ contains derivative
operators as well, however, we have to replace \eqref{eq:f1Ratio1} and
\eqref{eq:f1Ratiotau} by 
\begin{align}
  \label{eq:f1RatioQE1}
  &\tt_1(1)f_{\be,\pbe }[z,\pz]\;\tt_1(1)^{-1}\,e^R
  \nonumber\\[0.3\baselineskip]
  &\ =\tt_1(1) f_{\be,\pbe }[z,\pz]\,e^R
  =e^{\i\phi_1} f_{\be,\pbe }[z,\pz]\,e^R,
  \\[0.8\baselineskip]\label{eq:f1RatioQEtau} 
  &\tt_1(\tau)f_{\be,\pbe }[z,\pz]\;\tt_1(\tau)^{-1}\,
  e^{\i\pi N_\phi (2z_1+\tau)}\,e^R
  \nonumber\\[0.3\baselineskip]
  &\ =\tt_1(\tau)\,f_{\be,\pbe }[z,\pz]\, e^R
  =e^{\i\phi_\tau}\,f_{\be,\pbe }[z,\pz]\, e^R,
\end{align}
where $\tt_1(\xi)$ is according to \eqref{eq:ttilde} given by 
\begin{align}
  \label{eq:t1tilde}
  \tt_1(\xi) 
  &=\ex{-\frac{1}{4l^2}\bar\xi (2z_1+\xi)}\ext{\xi\partial_{z_1}}.
\end{align}
$\tt_1(\xi)$ translates both $z_1\to z_1+\xi$ and $2l^2\partial_{z_1}\to
2l^2\partial_{z_1}+\bar\xi$.  (Some care is required because the 
operators effecting these actions do not commute.)

Regarding the quasi-periodicity in the quasi-electron coordinate
$\be$, we introduce yet another set of ladder operators for the
quasi-electron coordinate,
\begin{align}
  \label{eq:bcLadder}
  \bar c&=\sqrt{2m}l\hspace{2pt} e^{-S}\hspace{1pt}\partial_\be e^{+S},
   \\[0.5\baselineskip] \label{eq:bcdagLadder}
  \bar c^\dagger&=\sqrt{2m}l\hspace{2pt} 
   e^{-S}\hspace{-1pt} 
   \left(-\partial_{\eta}+\frac{1}{2ml^2}\bar\eta\right) e^{+S},
\end{align}
where $S$ is still given by \eqref{eq:Seta[z]}.  (Again, for the
purposes of defining the operators, we could have just as well taken
$S=|\eta|^2/4ml^2$, as the terms depending on the $z_i$'s cancel in
\eqref{eq:bcLadder} and \eqref{eq:bcdagLadder}.)  $\bar c$ and $\bar
c^\dagger$ obey the commutation relation $\comm{\bar c}{\bar
  c^\dagger}=1$ and trivially commute with $a$, $a^\dagger$, $b$, and
$b^\dagger$.  We define the magnetic translation operator for the
quasi-electron coordinate,
\begin{align}
  \label{eq:tbe}
  t_\be(\xi)
  \equiv\ex{\frac{1}{\sqrt{2m}l}(\bar\xi\bar c-\xi\bar c^\dagger)}.
\end{align}
As for the quasi-hole, $t_\be(\xi_1)$ and $t_\be(\xi_2)$ commute only
if the area spanned by $\xi_1$ and $\xi_2$ in the plane contains $m$
times an integer number $N_\phi$ of magnetic Dirac flux quanta.

As in the previous sections, we further introduce an operator
\begin{align}
  \label{eq:tbetilde}
  \tt_\be(\xi) &\equiv e^{+S} t_\be(\xi) e^{-S}
  \nonumber \\[0.2\baselineskip] 
  &=\ex{\bar\xi\partial_\be+\xi\partial_{\eta}
    -\frac{1}{2ml^2}\xi\bar\eta}
  \nonumber \\[0.2\baselineskip] 
  &=\ex{-\frac{1}{4ml^2}\xi (2\be+\bar\xi)}\ext{\bar\xi\partial_\be},
\end{align}
where we again omitted the term involving $\partial_{\eta}$ (knowing
that it will only act on functions of $\bar\eta$, not $\eta$).  Acting
on the factor $e^R$ to the left of $f_{\be,\pbe}[z,\pz]$ in
\eqref{eq:psibe[z]} yields
\begin{align}
  \label{eq:ttilde1ONe^bareta^2}
  \tt_\be(m)\, e^{{\be^2}\hspace{-2pt}/{4ml^2}}&=e^{{\be^2}\hspace{-2pt}/{4ml^2}},
  \\[0.3\baselineskip] \label{eq:ttildetauONe^bareta^2}
  \tt_\be(m\tau)\, e^{{\be^2}\hspace{-2pt}/{4ml^2}}&=
  e^{-i\pi N_\phi (2\be+m\bar\tau)}\,e^{{\be^2}\hspace{-2pt}/{4ml^2}}.
\end{align}
In analogy to \eqref{eq:tetaPBC},
we expect PBCs for the quasi-electron according to
\begin{align}
\label{eq:bartetaPBC}
  t_\be(m\xi_\a)\hspace{1pt}\psi_\be[z]
  &=e^{-\i\phi_\a}(-1)^{m-1}\hspace{1pt}\psi_\be[z],
  \hspace{5pt}\text{for}\ \a=1,\tau,
\end{align}
where $\xi_1=1$, $\xi_\tau=\tau$.  
Note that we have reversed the signs of the boundary phases $\phi_1$
and $\phi_\tau$ as compared to the quasi-hole, accounting for the
quasi-electron being the quasi-holes antiparticle.  These boundary
conditions translate into the following conditions for $f_{\be,\pbe
}[z,\pz]$,
\begin{align}
  \label{eq:f1RatioQEm}
  \tt_\be(m) &\,f_{\be,\pbe }[z,\pz]\;\tt_\be(m)^{-1}\,e^R
  \nonumber\\[0.3\baselineskip]
  &=\tt_\be(m)\,f_{\be,\pbe }[z,\pz]\,e^R
  \nonumber\\[0.3\baselineskip]
  &=e^{-\i\phi_1}(-1)^{m-1} f_{\be,\pbe }[z,\pz]\,e^R,
  \\[0.8\baselineskip]\label{eq:f1RatioQEmtau} 
  \tt_\be(m\tau) &\,f_{\be,\pbe }[z,\pz]\;\tt_\be(m\tau)^{-1}
  e^{-\i\pi N_\phi (2\be+m\bar\tau)} \,e^R
  \nonumber\\[0.3\baselineskip]
  &=\tt_\be(m\tau)\,f_{\be,\pbe }[z,\pz]\, e^R
  \nonumber\\[0.3\baselineskip]
  &=e^{-\i\phi_\tau}(-1)^{m-1}\,f_{\be,\pbe }[z,\pz]\, e^R.
\end{align}
The challenge to identify the quasi-electron wave function is now
reduced to constructing a quasi-electron (wave) function $f_{\be,\pbe
}[z,\pz]$ with transformation properties
\eqref{eq:f1RatioQE1}, \eqref{eq:f1RatioQEtau} and
\eqref{eq:f1RatioQEm}, \eqref{eq:f1RatioQEmtau} with $\tt_{1}(\xi)$ and
$\tt_\be(\xi)$ given by \eqref{eq:t1tilde}
and \eqref{eq:tbetilde}, respectively.

From \eqref{eq:deffeta}, \eqref{eq:qeFactor}, \eqref{eq:FSingle},
and \eqref{eq:Feta}, we expect $f_{\be,\pbe}[z,\pz]$ to contain
factors
\begin{align}
  \label{eq:defQ}
  Q_\be[\pz]&=\prod_{i=1}^N\thembtaubig{2l^2\partial_{z_i}-\be},
  \\[0.1\baselineskip] \label{eq:defJ}
  J[z]&=\prod_{i<j}^N\thetau{z_i-z_j}^m,
\end{align}
and
\begin{align}
  \label{eq:FpartialbeSingle}
  F_{\pbe}(Z)=\,&\exBig{\i K\left(Z-2l^2\pbe\right)} 
  \nonumber\\[0.2\baselineskip]  
  \cdot\,
  &\prod_{\nu=1}^{m} \thetaubig{Z-{2l^2\pbe}-Z_\nu},
\end{align}
where $K$ is a real parameter $|K|\le\pi m$, the center-of-mass
coordinate now includes the quasi-electron coordinate in form of a
partial derivative, and all the zeros $Z_\nu$ are located in the
principal region.  (The derivative in $\be$ in
\eqref{eq:FpartialbeSingle} is required because $\tt_\be(\xi)$
transforms $\be\to\be+\bar\xi$ but $2ml^2\pbe\to2ml^2\pbe+\xi$.)

Since the transformation properties of these factors under
translations of $z_1$ by $1$ or $\be$ by $m$ are rather trivial,
we proceed by investigating translations of $z_1$ by $\tau$ or 
$\be$ by $m\bar\tau$.

Explicit evaluation of the action of $\tt_{1}(\tau)$ on
\eqref{eq:defQ}, \eqref{eq:FpartialbeSingle}, and \eqref{eq:defJ}
yields
\begin{align}
  \label{eq:Qtau} 
  \tt_1(\tau)\,&Q_\be[\pz]\;\tt_1(\tau)^{-1}
  \nonumber\\[0.3\baselineskip]  
  =\,&Q_\be[\pz]\,(-1)\, e^{\i\pi\bar\tau} 
  \exBig{2\pi\i\left(2l^2\partial_{z_1}-\be\right)},
  \\[0.7\baselineskip]  
  \label{eq:Ftau} 
  \tt_1(\tau)\,&F_{\pbe}(Z)\;\tt_1(\tau)^{-1}
  \nonumber\\[0.3\baselineskip]  
  =\,&F_{\pbe}(Z)\,(-1)^{m}\, e^{\i K\tau}\,e^{-\i\pi\tau m} 
  \nonumber\\[0.3\baselineskip]  
  \cdot\,&\ex{\!-2\pi\i \left[mZ - 2ml^2\pbe -\sum_{\nu=1}^m Z_\nu\right]},
  \\[0.7\baselineskip]  
  \label{eq:Jtau} 
  \tt_1(\tau)\,&J[z]\;\tt_1(\tau)^{-1}
  \nonumber\\[0.3\baselineskip]  
  =\, &J[z]\,(-1)^{m(N-1)}\, e^{-\i\pi\tau m(N-1) }\,e^{-2\pi\i m (N z_1-Z)}.  
\end{align}
While there are the usual cancelations between the terms on the right
in \eqref{eq:Ftau} and \eqref{eq:Jtau}, there is a complete mismatch
with the terms in \eqref{eq:Qtau}.  To match them, we need to generate
an additional factor
\begin{align}
  \nonumber
  \exBig{-2\pi\i\left(2l^2\partial_{z_1}-\be\right)}
  \exBig{2\pi\i \left(z_1 - 2ml^2\pbe\right)}
  e^{\i\pi(\tau-\bar\tau)}
\end{align}
between \eqref{eq:Qtau} and \eqref{eq:Ftau} under translation by
$\tt_{1}(\tau)$.  To this end, we introduce the Gaussian
\begin{align}
  \label{eq:G}
  &G_{\be,\pbe}[z,\pz]=\prod_{i=1}^N \ex{\frac{1}{4ml^2N} D_i^2},
\end{align}
where
\begin{align}
  \label{eq:Di}
  D_i\equiv z_i-2l^2\partial_{z_i}+\be-2ml^2\pbe.
\end{align}
The transformation properties of $G_{\be,\pbe}[z,\pz]$ are (see
Appendix)
\newlength\shift\setlength\shift{-20pt}
\begin{align}
  \label{eq:G1}
  \tt_1(1)\hspace{-\shift}&\hspace{\shift}
  \,G_{\be,\pbe}[z,\pz] \;\tt_1(1)^{-1}=\,G_{\be,\pbe}[z,\pz],
  \\[0.7\baselineskip]  
  \label{eq:Gtau} 
  \tt_1(\tau) \hspace{-\shift}&\hspace{\shift}
  \,G_{\be,\pbe}[z,\pz] \;\tt_1(\tau)^{-1}
  \nonumber\\[0.3\baselineskip]  
  =\,& 
  \exBig{-2\pi\i\left(2l^2\partial_{z_1}-\be\right)}
  \nonumber\\[0.3\baselineskip] \cdot\,&
  \,G_{\be,\pbe}[z,\pz] 
  \ex{\!-\frac{2\pi\i}{N}D\!}
  \nonumber\\[0.3\baselineskip]  
  \cdot\,&\exBig{2\pi\i \left(z_1 - 2ml^2\pbe\right)}\, e^{\i\pi(\tau-\bar\tau)},
\end{align}
where
\begin{align}
  \label{eq:D}
  D\equiv\sum_{i=1}^N D_i.
\end{align}
Combining these properties, we find that
\begin{align}
  \label{eq:fQE}
  f_{\be,\pbe}[z,\pz]&=
  Q_\be[\pz]\,G_{\be,\pbe}[z,\pz]\,F_{\pbe}(Z)\,J[z]
\end{align}
transforms as 
\begin{align}
  \label{eq:T1fT1}
  \tt_1(\tau)\,&f_{\be,\pbe}[z,\pz]\;\tt_1(\tau)^{-1}
  \nonumber\\[0.3\baselineskip]  
  =\;&Q_\be[\pz]\,G_{\be,\pbe}[z,\pz]\,
  \ex{\!-\frac{2\pi\i}{N}D\!}
  F_{\pbe}(Z)\,J[z]
  \nonumber\\[0.3\baselineskip]  
  \cdot\;&(-1)^{N_\phi}e^{\i K\tau}\,\ex{2\pi\i\sum_{\nu=1}^{m}Z_\nu}
  e^{-\i\pi N_\phi (2z_1+\tau)}.
\end{align}
With
\begin{align}
  \label{eq:Dcomm}
  \comm{D}{F_{\pbe}(Z)}=\bigcomm{D}{J[z]}=0
  \hspace{10pt}\text{and}\hspace{10pt} D_i\, e^R=0,
\end{align}
we find with \eqref{eq:T1fT1} that \eqref{eq:fQE} satisfies
\eqref{eq:f1RatioQE1} and \eqref{eq:f1RatioQEtau} provided that the
boundary conditions \eqref{eq:Fbc1} and \eqref{eq:Fbctau} are
satisfied.

\setlength\shift{0pt}
Under translations of the quasi-electron coordinate $\be$ by
$m\bar\tau$ we find that the 
factors $Q_\be[\pz]$, $F_{\pbe}(Z)$, and $J[z]$ transform as
\begin{align}
  \label{eq:Qmtau} 
  \tt_\be(m\tau)\,\hspace{-\shift}&\hspace{\shift}
  Q_\be[\pz]\;\tt_\be(m\tau)^{-1}
  \nonumber\\[0.3\baselineskip]  
  =\,&Q_\be[\pz]\,(-1)^{mN}\, e^{\i\pi\bar\tau m^2 N} 
  \nonumber\\[0.3\baselineskip]  
  \cdot\,&\prod_{i=1}^N
  \exBig{\!-2\pi\i m\!\left(2l^2\partial_{z_i}-\be\right)},
  \\[0.7\baselineskip]  
  \label{eq:Fmtau} 
  \tt_\be(m\tau)\,\hspace{-\shift}&\hspace{\shift}
  F_{\pbe}(Z)\;\tt_\be(m\tau)^{-1}
  \nonumber\\[0.3\baselineskip]  
  =\,&F_{\pbe}(Z)\,(-1)^{m}\, e^{-\i K\tau}\,e^{-\i\pi\tau m} 
  \nonumber\\[0.3\baselineskip]  
  \cdot\, &\ex{2\pi\i \left[mZ - 2ml^2\pbe -\sum_{\nu=1}^m Z_\nu\right]},
  \\[0.7\baselineskip]  
  \label{eq:Jmtau} 
  \tt_\be(m\tau)\,&J[z]\;\tt_\be(m\tau)^{-1}=J[z],
\end{align}
while the Gaussian factor transforms as (see Appendix)
\begin{align}
  \label{eq:Gm}
  \tt_\be(m)\,\hspace{-\shift}&\hspace{\shift}
  G_{\be,\pbe}[z,\pz] \;\tt_\be(m)^{-1}=\,G_{\be,\pbe}[z,\pz],
  \\[0.7\baselineskip]  
  \label{eq:Gmtau} 
  \tt_\be(m\tau)\,\hspace{-\shift}&\hspace{\shift}
  G_{\be,\pbe}[z,\pz] \;\tt_\be(m\tau)^{-1}
  \nonumber\\[0.1\baselineskip]  
  =\,& \prod_{i=1}^N\exBig{2\pi\i m\left(2l^2\partial_{z_i}-\be\right)}
  \nonumber\\[0.1\baselineskip]  
  \cdot\,&\,G_{\be,\pbe}[z,\pz]\,\exbig{2\pi\i m D}
  \nonumber\\[0.1\baselineskip]  
  \cdot\, &\prod_{i=1}^N\exBig{\!-2\pi\i m\left(z_i - 2ml^2\pbe\right)}
  \,e^{\i\pi m^2 N(\tau-\bar\tau)}.
  \nonumber\\[-0.7\baselineskip]
\end{align}
Taken together, we obtain the transformation properties 
\begin{align}
  \label{eq:fQEmtau}
  \tt_\be(m\tau)\,&f_{\be,\pbe}[z,\pz]\;\tt_\be(m\tau)^{-1}
  \nonumber\\[0.5\baselineskip]  
  =\;&f_{\be,\pbe}[z,\pz]\, 
  \nonumber\\[0\baselineskip]  
  \cdot\, & (-1)^{N_\phi}\,  e^{-\i K\tau}  
  \ex{\!-2\pi\i\sum_{\nu=1}^m Z_\nu} (-1)^{m-1}
  \nonumber\\[0.3\baselineskip]  
  \cdot\, & \exbig{2\pi\i m D}\, 
  e^{\i\pi\tau m N_\phi}\ex{2\pi\i N_\phi\, 2ml^2\pbe }.
  \nonumber\\[0\baselineskip]  
\end{align}
The last term generated in \eqref{eq:fQEmtau} does not resemble the
term required by \eqref{eq:f1RatioQEmtau}, but since
\begin{align}
  \hspace{\shift}&\hspace{-\shift}
  e^{\i\pi\tau m N_\phi} \ex{2\pi\i N_\phi\, 2ml^2\pbe }\,
  e^{-i\pi N_\phi (2\be+m\bar\tau)}\,e^{{\be^2}\hspace{-2pt}/{4ml^2}}
  \nonumber\\[0.3\baselineskip]  
  &=\exbig{m(\tau-\bar\tau)\pbe}\,
  e^{{\be^2}\hspace{-2pt}/{4ml^2}}\, e^{-i\pi N_\phi(2\be-m(\tau-\bar\tau)) }
  \nonumber\\[0.3\baselineskip]  
  &=  e^{{\be^2}\hspace{-2pt}/{4ml^2}},
\end{align}
we find with \eqref{eq:fQEmtau} and \eqref{eq:Dcomm} that
\eqref{eq:fQE} satisfies \eqref{eq:f1RatioQEm} and
\eqref{eq:f1RatioQEmtau} provided that the boundary conditions
\eqref{eq:Fbc1} and \eqref{eq:Fbctau} are satisfied.

A quasi-electron wave function with the correct quasi-periodicity
under magnetic translations is hence given by \eqref{eq:psibe[z]} with
\eqref{eq:Seta[z]}, \eqref{eq:fQE} and \eqref{eq:R}.

The final form, however, may be simplified significantly.  Writing out
\eqref{eq:fQE} in full, we have
\setlength\shift{15pt}
\begin{align}
  \label{eq:fAllTerms}
  f_{\be,\pbe}[z,\pz] 
  =&\prod_{i=1}^N\thembtaubig{2l^2\partial_{z_i}-\be}
  \nonumber\\[0.1\baselineskip]
  \cdot&\prod_{i=1}^N
  \ex{\frac{\left(z_i-2l^2\partial_{z_i}+\be-2ml^2\pbe\right)^2}{4ml^2N}}
  \nonumber\\[0.1\baselineskip]
  \cdot&\exBig{\i K\left(Z-2l^2\pbe\right)}
  \nonumber\\[0.1\baselineskip]
  \cdot&\prod_{\nu=1}^{m} \thetaubig{Z-{2l^2\pbe}-Z_\nu}
  \nonumber\\[0.1\baselineskip]
  \cdot&\prod_{i<j}^N\thetau{z_i-z_j}^m.
\end{align}
Note first that the argument of the Gaussian and the argument in
the line below commute,
\begin{align}
  \label{eq:GFcommute}
  \comm{z_i-2l^2\partial_{z_i}+\be-2ml^2\pbe}{Z-2l^2\pbe}=0,
\end{align}
which implies that we can interchange the orders of 
$G_{\be,\pbe}[z,\pz]$ and $F_{\pbe}(Z)$ in \eqref{eq:fAllTerms}.
Second, since
$Q_\be[\pz]$ and $F_{\pbe}(Z)$ commute, we may interchange them and
thereby bring the derivatives in the $z_i$ closer to the Jastrow
factor.  We may hence write
\begin{align}
  \label{eq:fAllTermsSimplified}
  f_{\be,\pbe}[z,\pz] 
  =&\exBig{\i K\left(Z-2l^2\pbe\right)}
  \nonumber\\[0.1\baselineskip]
  \cdot&\prod_{\nu=1}^{m} \thetaubig{Z-{2l^2\pbe}-Z_\nu}
  \nonumber\\[0.1\baselineskip]
  \cdot&  \prod_{i=1}^N\thembtaubig{2l^2\partial_{z_i}-\be}
  \nonumber\\[0.1\baselineskip]
  \cdot&\prod_{i=1}^N
  \ex{\frac{\left(z_i-2l^2\partial_{z_i}+\be-2ml^2\pbe\right)^2}{4ml^2N}}
  \nonumber\\[0.1\baselineskip]
  \cdot&
  \prod_{i<j}^N\thetau{z_i-z_j}^m.
\end{align}

Substitution of \eqref{eq:fAllTermsSimplified} into
\eqref{eq:psibe[z]} with \eqref{eq:Seta[z]} and \eqref{eq:R} allows
now for significant simpification.  Since $f_{\be,\pbe}[z,\pz]$ acts
on $e^R$, and
\begin{align}
  \left(\be-2ml^2\pbe\right)\, e^{{\be^2}\hspace{-2pt}/{4ml^2}}&=0,
  \nonumber\\[0.3\baselineskip] \nonumber
  \left(z_i-2l^2\partial_{z_i}\right)\,
  e^{{z_i^2}\hspace{-2pt}/{4l^2}} J[z]
  &= e^{{z_i^2}\hspace{-2pt}/{4l^2}}
  \left(-2l^2\partial_{z_i}\right) J[z],\nonumber
\end{align}
we obtain the final form of the quasi-electron wave function,
\setlength\shift{20pt}
{
\begin{align}
  \label{eq:QEfinal}
  \psi_\be[z]=&\, 
  e^{-S}\,e^{\i K\left(Z-2l^2\pbe\right)}
  \nonumber\\[0.1\baselineskip]
  \cdot&\prod_{\nu=1}^{m} \thetaubig{Z-{2l^2\pbe}-Z_\nu}
  \,\cdot\,e^{{\be^2}/{4ml^2}}
  \nonumber\\[0.1\baselineskip]
  \cdot&\prod_{i=1}^N\thembtaubig{2l^2\partial_{z_i}-\be}\cdot 
  \prod_{i=1}^N e^{{z_i^2}/{4l^2}}
  \nonumber\\[0.3\baselineskip]
  \cdot&\prod_{i=1}^N
  \ex{\frac{l^2\partial_{z_i}^2}{mN}}\cdot 
  \prod_{i<j}^N\thetau{z_i-z_j}^m.
\end{align}
where $S$ is given by \eqref{eq:Seta[z]}.  Since we have not found a
plausible way to motivate the form \eqref{eq:QEfinal} directly,
however, we present the results in the order we have obtained them.
An additional advantage of \eqref{eq:QEfinal} over \eqref{eq:psibe[z]}
with \eqref{eq:fAllTerms} and \eqref{eq:R} is that the generalization
to several quasi-electrons is straightforward.
 
This concludes our derivation of the Laughlin's quasi-electron states
on the torus.

A very valid question to ask at this point is how feasible it is to
evaluate \eqref{eq:QEfinal} explicitly for applications in numerical
studies.  To begin with, the odd Jacobi theta functions in
\eqref{eq:QEfinal} are according to \eqref{eq:thetadef} defined as
\begin{align}
  \label{eq:theta1def}
  \thetau{z}=\sum_{n\in\mathbb{Z}+\frac{1}{2}} 
  \,e^{\pi\i n^2\tau}\,e^{\pi\i n}\,e^{2\pi\i n z},
\end{align}
where we have adjusted $n$ such that we sum only over half-integer
values.  If we write $\tau=\tau'+\i\tau''$, with
$\tau',\tau''\in\mathbb{R}$, $\tau''>0$, we see that the sum
\eqref{eq:theta1def} converges the quicker the larger $\tau''$.  Since
the quantum Hall wave functions are modular invariant, with modular
transformations generated by
\begin{align}
  \label{eq:modtransf}
  \text{T: }\tau\to\tau+1\quad 
  \text{and\quad S: }\tau\to-\frac{1}{\tau},
\end{align}
we can always choose a unit cell such that
$\tau''\ge\frac{1}{2}\sqrt{3}$.  For the more typical rectangular unit
cells we may even choose $\tau''\ge 1$.  Even if we assume that
$\text{Im(z)}=-\tau''$ in \eqref{eq:theta1def}, we see that the series
decays as $e^{-\pi (n^2-2n)\tau''}$, which for $\tau''=1$ is of order
1 for $n=\frac{1}{2}$ and of order $10^{-27}$ for $n=\frac{11}{2}$,
while the sum of all terms for larger $|n|$ is negligible by
comparison.  It hence appears sufficient to keep the 12 terms from
$n=-\frac{11}{2}$ to $+\frac{11}{2}$.  The derivatives in the Jacobi
theta functions amount hence only to derivative operators in 12 terms
of a sum, which in principle can be evaluated using programs
performing symbolic manipulations of mathematical expressions.

A remaining obstacle is posed by the combination of derivatives in the
theta functions with the Gaussians in $\be^2$ and $z_i^2$ in the
second and third line of \eqref{eq:QEfinal}, respectively, or the
Gaussian in the derivatives acting on the arguments of the theta
functions in the last line.  Using the Baker--Hausdorff formula
\begin{align}
  \nonumber
  e^x\, e^y = e^{x+y+\frac{1}{2}\comm{x}{y}
    +\frac{1}{12}\comm{x}{\comm{x}{y}}
    +\frac{1}{12}\comm{\comm{x}{y}}{y}},
\end{align}
which truncates as stated if both $\comm{x}{\comm{x}{y}}$ and
$\comm{\comm{x}{y}}{y}$ commute with both $x$ and $y$, we may write
\begin{align}
  \ex{a\partial_z^2} e^{2\pi\i n z} 
  &= e^{2\pi\i n (z+2 a\partial_z)} \ex{a\partial_z^2},
  \nonumber\\[0.5\baselineskip]
  e^{2\pi\i n\,(2l^2\partial_z)}\, e^{{z^2}/{4l^2}}
  &=  e^{{z^2}/{4l^2}}\, e^{2\pi\i n\,(2l^2\partial_z + z)},
  \nonumber\\[0.5\baselineskip]
  e^{2\pi\i n\,(-2l^2\pbe)}\, e^{{\be^2\!}/{4ml^2}}
  &=  e^{{\be^2\!}/{4ml^2}}\, e^{2\pi\i n\,(-2l^2\pbe -\be/m)}.
  \nonumber
\end{align}
Commuting the Gaussian factors through the theta functions will hence
only generate terms inside the theta functions which are linear in the
arguments of the Gaussian.  We may use these commutators to rewrite
\eqref{eq:QEfinal} as
\begin{align}
  \label{eq:QEfinal2}
  \psi_\be[z]=\, & e^{-S}\,e^{{\be^2}/{4ml^2}}\,
  \prod_{i=1}^N e^{{z_i^2}/{4l^2}}\cdot e^{\i K\left(Z-2l^2\pbe-\be/m\right)}
  \nonumber\\[0.1\baselineskip]
  \cdot\,&
  \prod_{\nu=1}^{m} \thetaubigg{Z-{2l^2\pbe-\frac{\be}{m}}-Z_\nu}
  \nonumber\\[0.1\baselineskip]
  \cdot\,&
  \prod_{i=1}^N\thembtaubig{2l^2\partial_{z_i}+z_i-\be}
  \nonumber\\[0.3\baselineskip]
  \cdot\,&\prod_{i<j}^N\thetaubigg{z_i-z_j+\frac{2l^2(
      \partial_{z_i}-\partial_{z_j})}{mN}}^m.
\end{align}
The evaluation of this expression still constitutes a challenge, as
each theta function represents a sum of about 12 terms with the
argument in exponential functions.  Since the arguments now consist of
linear functions in the coordinates and derivatives taken in these
coordinates, however, the combinations may now be evaluated by
repeated application of the Baker--Hausdorff formula, where each
commutator gives a contribution which does not depend on the
coordinates (but only of the indices $n$ of the sums
\eqref{eq:theta1def}).  Even though we believe that the form
\eqref{eq:QEfinal2} is useful for the explicit evaluation of the
quasi-electron wave function, we prefer to consider the form
\eqref{eq:QEfinal} as the final result of our derivation.  This form 
displays the desired transformation properties more succinctly.}   

\section{Conclusion}
\label{sec:conclusion}

In this article, we set up an operator formalism for Landau levels and
magnetic translations, and used it to formulate Laughlin's wave
functions for fractionally quantized Hall states subject to PBCs.  The
results have been known for three decades for the ground states
and the quasi-hole excitations.  They are, however, original for the
technically more challenging quasi-electron excitations, which were
left as an open problem in the classic work by Haldane and
Rezayi~\cite{haldane-85prb2529}.  Comparing the final form
\eqref{eq:QEfinal} 
\begin{align}
  \psi_\be[z]=&\, 
  e^{-S}\,e^{\i K\left(Z-2l^2\pbe\right)}
  \nonumber\\[0.1\baselineskip]
  \cdot&\prod_{\nu=1}^{m} \thetaubig{Z-{2l^2\pbe}-Z_\nu}
  \cdot e^{{\be^2}/{4ml^2}}
  \nonumber\\[0.1\baselineskip]
  \cdot&\prod_{i=1}^N\thembtaubig{2l^2\partial_{z_i}-\be}\cdot
  \prod_{i=1}^N e^{{z_i^2}/{4l^2}}
  \nonumber\\[0.3\baselineskip]\nonumber
  \cdot&\prod_{i=1}^N
  \ex{\frac{l^2\partial_{z_i}^2}{mN}}\cdot
  \prod_{i<j}^N\thetau{z_i-z_j}^m
\end{align}
of the quasi-electron wave function with the final form of the
quasi-hole wave function
\begin{align}
  \label{eq:QHfinal}
  \psi_\eta[z]=&\, 
  e^{-S}\,e^{\i K\left(Z+\eta/m\right)}
  \nonumber\\[0.1\baselineskip]
  \cdot&\prod_{\nu=1}^{m}\thetaubigg{Z+\frac{\eta}{m}-Z_\nu}
  \cdot e^{{\eta^2}/{4ml^2}}
  \nonumber\\[0.1\baselineskip]
  \cdot&\prod_{i=1}^N\thetau{z_i-\eta}\cdot
  \prod_{i=1}^N e^{{z_i^2}/{4l^2}}
  \nonumber\\[0.1\baselineskip]
  \cdot&\prod_{i<j}^N\thetau{z_i-z_j}^m,
\end{align}
we see that the quasi-electron function is very close to what we would
have expected from the known wave functions in the plane, which
already contains the derivative operators in the coordinates $z_i$ and
$\be$ rather than the quasi-electron coordinate $\eta$ in the complex
plane.  That we need to subtract $2l^2\pbe$ rather than $\be/m$ or
even $\eta/m$ from the center-of-mass coordinate $Z$ is dictated by
the quasi-periodic boundary conditions and the requirement that
$\psi_\be[z]$ is given by $e^{-S}$ times an analytic function in both
$z$ and $\be$.  The only part we would not have been able to
anticipate is the Gaussian factor in the last line of
\eqref{eq:QEfinal}.  This is a minor modification, which nonetheless
required a major effort to obtain.

The final form \eqref{eq:QEfinal}, however, reveals something else we
could not have easily anticipated.  Since we cannot commute the
factors $e^{{z_i^2}/{4l^2}}$ through the quasi-electron theta function
to the right of them without modifying their arguments as indicated in
\eqref{eq:QEfinal2}, we see that a formulation of the quasi-electron
in the otherwise more compact Landau gauge used by Haldane and
Rezayi~\cite{haldane-85prb2529} would require factors
$e^{-{z_i^2}/{4l^2}}$ and $e^{{z_i^2}/{4l^2}}$ to the left and to the
right of this theta function, respectively.  This indicates that
obtaining the quasi-electron in this gauge would presumably be
significantly more difficult, which may explain why the problem has
not been solved previously.

We have further seen that the quasi-electron wave function we have
derived for PBCs in this article does not lend itself well to explicit
evaluation, which sincerely limits the applicability and usefulness of
our result.  To find out about these limitations, however, it was
necessary to obtain the wave function in the first place.  The main
result of our study may hence be that it is possible to generalize
Laughlin's quasi-electron excitation to PBCs, but that it is not
practical to work with the ensuing form.

\begin{acknowledgments}
  This work was supported by the ERC starters grant TOPOLECTRICS under
  ERC-StG-Thomale-336012.
\end{acknowledgments}

\appendix* 
\section{Transformation properties of $G_{\be,\pbe}[z,\pz]$}
In this appendix, we will sketch the derivation of \eqref{eq:Gtau} and
\eqref{eq:Gmtau}.  To begin with, we introduce the shorthand notation
\begin{align}
  \label{eq:Gi}
  G_{\be,\pbe}[z,\pz]=G=\prod_{i=1}^N G_i,\quad G_i=\ex{\frac{D_i^2}{4ml^2N}}
\end{align}
with 
$D_i=A_i+B_i$, $A_i=\be-2l^2\partial_{z_i}$, $B_i=z_i-2ml^2\pbe$.
We then use 
\begin{align}
  \label{eq:Tau-Taubar}
  \tau-\bar\tau= 2\i\,2\pi l^2 N_\phi=2\pi\i\, 2l^2\, (Nm-1)
\end{align}
to evaluate 
\setlength\shift{6pt}
\begin{align}
  \label{eq:Gin}
   \tt_i(n\tau) \hspace{-\shift}&\hspace{\shift}\,G_i\;\tt_i(n\tau)^{-1}
  \nonumber\\[0.3\baselineskip]  
  =\,&\ex{\frac{1}{4ml^2N} \Bigl(
      \left(D_i-2\pi\i n\,2l^2\right)+2\pi\i n\,2l^2 mN\Bigr)^2} 
  \nonumber\\[0.3\baselineskip]  
  =\,&e^{2\pi\i n A_i}\, e^{2\pi\i n B_i}\,
  \nonumber\\[0.3\baselineskip]  
  \cdot\,&\ex{\frac{1}{4ml^2N} \left(D_i-2\pi\i n\,2l^2\right)^2}
  \,e^{(2\pi\i n l)^2 (Nm-m-1)}
  \nonumber\\[0.3\baselineskip]  
  =\,&e^{2\pi\i n A_i}\, e^{2\pi\i n\,(-2l^2 m\pbe)}\,
  G_i\,e^{2\pi\i n z_i}\,e^{(2\pi\i n l)^2 (Nm-m-1)}.
\end{align}
We proceed
\begin{align}
  \label{eq:Gn}
   \tt_i(n\tau) \hspace{-\shift}&\hspace{\shift}\,G\;\tt_i(n\tau)^{-1}
   =\tt_i(n\tau)\,G_i\;\tt_i(n\tau)^{-1} \prod_{j=1\atop (j\ne i)}^N G_j
  \nonumber\\[0.3\baselineskip]  
  =\,&e^{2\pi\i n A_i}\, e^{2\pi\i n\,(-2l^2 m\pbe)}\,
  G\,e^{2\pi\i n z_i}\,e^{(2\pi\i n l)^2 (Nm-m-1)}
  \nonumber\\[0.3\baselineskip]  
  =\,&e^{2\pi\i n A_i}\, 
  \prod_{j=1}^N\ex{\frac{1}{4ml^2N} \left(D_j-2\pi\i n\,2l^2 m\right)^2}
  \nonumber\\[0.3\baselineskip]  
  \cdot\,&\,e^{2\pi\i n B_i}\, e^{(2\pi\i n l)^2 (Nm-m-1)}
  \nonumber\\[0.3\baselineskip]  
  =\,&e^{2\pi\i n A_i}\, G\,e^{-2\pi\i n D/N}\,
  e^{2\pi\i n B_i}\,e^{\pi\i n^2 (\tau-\bar\tau)},
\end{align}
where $D$ is given by \eqref{eq:D}.
For $n=1$, this reduces to \eqref{eq:Gtau}.  To verify
\eqref{eq:Gmtau}, we write
\begin{align}
  \label{eq:GmtauVerify}
  \tt_\be(m\tau)\,\hspace{-\shift}&\hspace{\shift} G\;\tt_\be(m\tau)^{-1}
  =\prod_{i=1}^N\tt_i(-m\tau)\hspace{4pt}G\,\prod_{i=1}^N\tt_i(-m\tau)^{-1} 
  \nonumber\\[0.3\baselineskip]  
  =\,&\prod_{i=1}^N e^{-2\pi\i m A_i}\, G\, e^{2\pi\i m D}\,
  \prod_{i=1}^N e^{-2\pi\i m B_i}\;e^{\pi\i m^2 N(\tau-\bar\tau)},
\end{align}
where we have used 
\begin{align}
  \nonumber
  e^{-2\pi\i m B_i}\,\tt_i(-m\tau)\,e^{\a D}\,\tt_i(-m\tau)^{-1}
  =e^{\a D}\,e^{-2\pi\i m B_i}
\end{align}
in obtaining the last line. 

\vfill\eject\vfill 

\vfill\eject
\end{document}